---------------------------------------------------------------------------
############################## RevTex Copy #################################
---------------------------------------------------------------------------

\documentstyle[aps]{revtex}

\newcommand{\bel}[1]{\begin{equation}\label{#1}}
\newcommand{\be}{\begin{equation}}
\newcommand{\ee}{\end{equation}}

\newcommand{\beal}[1]{\begin{eqnarray}\label{#1}}
\newcommand{\bea}{\begin{eqnarray}}
\newcommand{\eea}{\end{eqnarray}}

\newcommand{\bean}{\begin{eqnarray*}}
\newcommand{\eean}{\end{eqnarray*}}

\newcommand{\ba}{\begin{array}}
\newcommand{\ea}{\end{array}}

\newcommand{\bab}{\begin{abstract}}
\newcommand{\eab}{\end{abstract}}

\newcommand{\bml}{\begin{mathletters}}
\newcommand{\eml}{\end{mathletters}}

\newcommand{\q}{\quad}
\newcommand{\qq}{\quad\quad}

\newcommand{\bfm}[1]{\mbox{\boldmath $#1$}}

\newcommand{\dv}{\partial}

\newcommand{\bad}[2]{\left( \begin{array}{c}{#1}\\{#2}\end{array} \right)} 
\newcommand{\bat}[3]{\left( \begin{array}{c}{#1}\\{#2}\\
{#3}\end{array} \right)} 
\newcommand{\baq}[4]{\left( \begin{array}{c}{#1}\\{#2}\\{#3}\\
{#4}\end{array} \right)}

\newcommand{\bamd}[4]{\left( \begin{array}{cc}{#1}&{#2}\\
{#3}&{#4}\end{array} \right)}
\newcommand{\bamt}[9]{\left( \begin{array}{ccc}{#1}&{#2}&{#3}\\{#4}&{#5}&{#6}\\
{#7}&{#8}&{#9}\end{array} \right)}

\newcommand{\bamq}[4]{\left( \begin{array}{cccc}{#1}&{#2}&{#3}&{#4}\\}
\newcommand{\bamc}[5]{\left( \begin{array}{ccccc}{#1}&{#2}&{#3}&{#4}&{#5}\\}
\newcommand{\eam}{\end{array} \right)}

\newcommand{\raw}{\rightarrow}
\newcommand{\lraw}{\longrightarrow}

\newcommand{\lRaw}{\Longrightarrow}

\newcommand{\lLRw}{\Longleftrightarrow}

\newcommand{\ag}{\alpha}
\newcommand{\bg}{\beta}

\newcommand{\sg}{\sigma}

\newcommand{\bii}{\begin{itemize}}
\newcommand{\eii}{\end{itemize}}

\newcommand{\ben}{\begin{enumerate}}
\newcommand{\een}{\end{enumerate}}

\newcommand{\bq}{\begin{quote}}
\newcommand{\eq}{\end{quote}}

\newcommand{\bc}{\begin{center}}
\newcommand{\ec}{\end{center}}

\newcommand{\btb}{\begin{table}}
\newcommand{\etb}{\end{table}}

\newcommand{\bt}{\begin{tabular}}
\newcommand{\et}{\end{tabular}}

\newcommand{\br}{\begin{flushright}}
\newcommand{\er}{\end{flushright}}

\newcommand{\bl}{\begin{flushleft}}
\newcommand{\el}{\end{flushleft}}

\newcommand{\th}[1]{\thanks{#1}}

\newcommand{\f}[1]{\footnote{#1}}

\newcommand{\bref}{\begin{references}}
\newcommand{\eref}{\end{references}}

\newcommand{\bb}{\begin{thebibliography}{99}}
\newcommand{\eb}{\end{thebibliography}}
\newcommand{\bi}{\bibitem}

\newcommand{\btp}{\begin{titlepage}}
\newcommand{\etp}{\end{titlepage}}

\newcommand{\axp}[3]{Ann.~Phys.~(NY)                    {\bf #1},  #2  (19#3)}

\newcommand{\jxg}[3]{J.~Phys.~A                         {\bf #1},  #2  (19#3)}

\newcommand{\mxb}[3]{Mod.~Phys.~Lett.~A                 {\bf #1},  #2  (19#3)}

\newcommand{\nxd}[3]{Nuovo Cimento                      {\bf #1},  #2  (19#3)}

\newcommand{\pxf}[3]{Phys.~Rev.~D                       {\bf #1},  #2  (19#3)}

\newcommand{\pxxa}[3]{Prog.~Theor.~Phys.                {\bf #1},  #2  (19#3)}

\title{ODD DIMENSIONAL TRANSLATION BETWEEN\\ 
COMPLEX AND QUATERNIONIC QUANTUM MECHANICS}

\author{Stefano De Leo\th{Electronic mail: {\sl deleos@le.infn.it}}
 and Pietro Rotelli\th{Electronic mail: {\sl rotelli@le.infn.it}}}
\address{Dipartimento di Fisica, Universit\`a di Lecce\\
Istituto Nazionale di Fisica Nucleare, sezione di Lecce\\
Lecce, 73100, Italy}

\date{}

\draft

\begin{document}

\maketitle

\bab
We complete the rules of translation between standard complex quantum 
mechanics (CQM) and quaternionic quantum mechanics (QQM) with a complex 
geometry. In particular 
we describe how to reduce ($2n$+$1$)-dimensional complex matrices to 
{\em overlapping\/} ($n$+$1$)-dimensional quaternionic matrices with generalized 
quaternionic elements. This step resolves an outstanding difficulty with 
reduction of purely complex matrix groups within quaternionic QM and avoids 
{\em anomalous} eigenstates. As a result we present a more complete translation 
from CQM to QQM and viceversa.
\eab

\pacs{PACS numbers: 02.10.Tq, 02.20.Sv, 03.65.-w.\\
KeyWords: quaternions, group theory, quantum mechanics.}

\renewcommand{\thefootnote}{\sharp\arabic{footnote}}

\section{Introduction}
The possibility of using quaternions to describe standard quantum mechanics 
received a major thrust with the adoption by Horwitz and 
Bie\-denharn~\cite{hor} of a complex scalar product 
(complex geometry~\cite{rem}). These authors showed that this 
assumptions permits the 
definition (in analogy with standard theory) of tensor products between 
single particle wave functions without encountering intractable problems of 
interpretation and definition due to the non commuting multiplications of 
quaternionic wave functions\f{A minor criticism of the formalism of Horwitz 
and Biedenharn is that quaternions are used only for single particle wave 
functions in their paper. In practise they use the complex translation for 
multiparticle states, an unnecessary choice.}.

A second important step in this objective of translation was achieved 
with the introduction of the so-called {\em barred} operators~\cite{del1}, and 
specifically of $q\mid i$ ($q$-quaternion, $i$-one of the three imaginary 
units) which acts upon a (quaternionic) wave function 
$\psi$ by
\be 
(q\mid i) \psi = q\psi i \q. 
\ee
Originally these generalized quaternions were introduced to define an 
acceptable ($i$-complex hermitian) momentum operator 
($-\bfm{\dv} \mid i$) within a two-component quaternionic Dirac 
equation~\cite{rot}.

The distinction between left acting or right acting quaternions is of course 
justified by the non commutativity of quaternions. Instead of proceeding to 
consider the (possibly future) generalization to $1\mid j$ and $1\mid k$ 
terms, it was noted by the authors in a previous work~\cite{del2} 
that generalized 
quaternions i.~e. $q_{1}+q_{2}\mid i$ (indicated generically by 
${\cal H}\mid {\cal C}$) depend upon eight real numbers, the same degree of 
freedom of the most general two by two complex matrix. In fact the already 
well known identifications of $i$, $j$ and $k$ with 
$-\frac{i}{2}\bfm{\sg}$ (\bfm{\sg} the Pauli matrices) and of course 
$1$ (in $\cal H$) with the $2$-dimensional unit matrix can thus be extended to 
the most general $2$-dimensional complex matrix.

A particular set of rules for translation is given by
\bml
\beal{2a}
M=\bamd{c_{1}}{c_{2}}{c_{3}}{c_{4}} & \q \lRaw \q & M=q_{1} + q_{2} \mid i 
\qq [c_{1, \; ..., \; 4} \in {\cal C}(1, \; i) \; \; \mbox{and} \; \;
q_{1, \; 2} \in {\cal H}] \q ,
\eea
with
\bean
2q_{1} & = & c_{1}+c_{4}^{*}+j(c_{3}-c_{2}^{*}) \q ,\\
2iq_{2} & = & c_{1}-c_{4}^{*}-j(c_{3}+c_{2}^{*}) \q .
\eean
Equivalent to:
\bea
M=z_{1}+j\tilde{z}_{1} + (z_{2}+j\tilde{z}_{2}) \mid i  & \q \lRaw \q & 
M=\bamd{z_{1}+iz_{2}}{-(\tilde{z}_{1}^{*}+i\tilde{z}_{2}^{*})}
{\tilde{z}_{1}+i\tilde{z}_{2}}{z_{1}^{*}+iz_{2}^{*}} \q ,
\eea
\eml
This augments the so called symplectic rule for state vectors (column 
matrices)
\bml
\bea
\psi_{a} = \bad{z_{a}}{\tilde{z}_{a}} &  \q \lLRw  \q & \psi_{a}=z_{a}+
j\tilde{z}_{a} \q ,\\
\psi_{a}^{\dag} = \left( z_{a}^{*} \; \; \tilde{z}_{a}^{*} \right) &
\q \lLRw  \q & \psi_{a}^{\dag}=z_{a}^{*}-
j\tilde{z}_{a} \q ,
\eea
\eml
which in turn is consistent only within a complex geometry
\bea
\psi_{1}^{\dag} \psi_{2}=z_{1}^{*}z_{1}+z_{2}^{*}z_{2} &  \q \lLRw \q & 
(\psi_{1}^{\dag} \psi_{2})_{c}=\frac{1-i\mid i}{2} \; \psi_{1}^{\dag} \psi_{2}
\q .
\eea
We recall also that the rule for the tensor product of two state vectors 
$\psi_{1}$ and $\psi_{2}$ is thus automatic
\bea
\psi_{1} \otimes \psi_{2}=\baq{z_{1}z_{2}}{z_{1}\tilde{z}_{2}}
{\tilde{z}_{1}z_{2}}{\tilde{z}_{1}\tilde{z}_{2}} &  \q \lLRw \q & 
\psi_{1} \otimes \psi_{2}=\bad{z_{2}z_{1}+j
\tilde{z}_{2}z_{1}}{z_{2}\tilde{z}_{1}
+j\tilde{z}_{2}\tilde{z}_{1}}=
\bad{\psi_{2} z_{1}}{\psi_{2} \tilde{z}_{1}}
\q .
\eea
In the same way we may derive the rules for taking the tensor product of 
group elements, for example,
\bml
\be
g_{1}\otimes g_{2}=\bamd{c_{1}}{c_{2}}{c_{3}}{c_{4}} \otimes 
\bamd{z_{1}}{z_{2}}{z_{3}}{z_{4}}=\bamq{c_{1}z_{1}}{c_{1}z_{2}}{c_{2}z_{1}}
{c_{2}z_{2}} c_{1}z_{3} & c_{1}z_{4} & c_{2}z_{3} & c_{2}z_{4}\\
c_{3}z_{1} & c_{3}z_{2} & c_{4}z_{1} & c_{4}z_{2}\\
c_{3}z_{3} & c_{3}z_{4} & c_{4}z_{3} & c_{4}z_{3} \eam
\ee
can be translated as follows
\be
g_{1}\otimes g_{2}=(q_{1}+q_{2}\mid i)
\otimes  (p_{1}+p_{2}\mid i)=\bamd{\ag_{1}}{\ag_{2}}{\ag_{3}}{\ag_{4}}
(p_{1}+p_{2}\mid i) \q ,
\ee
\eml
where
\bean
\ag_{1} & = & (q_{1}+iq_{2})_{c} + (q_{1}+iq_{2})_{c}^{*} -i [
(q_{1}+iq_{2})_{c} - (q_{1}+iq_{2})_{c}^{*}]\mid i \q ,\\
\ag_{2} & = & (jq_{1}-kq_{2})_{c}^{*} + (jq_{1}-kq_{2})_{c} -i [
(jq_{1}-kq_{2})_{c}^{*} - (jq_{1}-kq_{2})_{c}]\mid i \q ,\\
\ag_{3} & = & -(jq_{1}+kq_{2})_{c} - (jq_{1}+kq_{2})_{c}^{*} +i [
(jq_{1}+kq_{2})_{c} - (jq_{1}+kq_{2})_{c}^{*}]\mid i \q ,\\
\ag_{4} & = & (q_{1}-iq_{2})_{c}^{*} + (q_{1}-iq_{2})_{c} -i [
(q_{1}-iq_{2})_{c}^{*} - (q_{1}-iq_{2})_{c}]\mid i \q .
\eean

This translation is only partial because it connects e.~g. $n$-dimensional 
quaternionic representations of quaternionic groups with {\em even} 
$2n$-dimensional 
complex representations of corresponding (isomorphic) complex groups and 
viceversa. 
Indeed we have presented elsewhere~\cite{del1} a comparison between the 
groups\f{With generalized 
quaternions we can also confront $U(2, \; c)$ with $U(1, \; q_{c})$, where 
$q_{c} \in {\cal H}\mid {\cal C}$.} 
$U(1, \; q)$ and $SU(2, \; c)$. Odd complex representations of the complex 
groups are excluded. There is also a problem with the irreducibility of odd 
dimensional complex representations of quaternionic groups (see below).

We note, in passing, that while matrix operators may contain generalized 
quaternions, the state vectors (wave functions in general) contain only 
quaternionic elements. This asymmetry correctly reproduces the {\em real} 
degrees of freedom between $n$-component complex column matrices and 
$n$-dimensional complex square matrices. This is because a quaternion has 4 
real degrees of freedom but a generalized quaternion 8.
\bc
\bt{|cc|c|ccc|}\hline
 & & & & & \\
~~~~~~Degrees of freedom on~~~~~~ &~~~$\cal R$~~~&~~~$\cal C$~~~&~~~$\cal H$~~~
& &~~~${\cal H}\mid {\cal C}$~~~\\ 
($n$ even) & & & & & \\ 
 & & & & & \\ \hline
 & & & & & \\
$\psi$ & $2n$ & $n$ & $n/2$ & & \\
 & & & & & \\ 
 & & $\uparrow \downarrow$ & & $\nwarrow \searrow$ & \\
 & & & & & \\
$M$ & $2n^{2}$ & $n^{2}$ & & & $(n/2)^{2}$ \\
 & & & & &\\ \hline
\et
\ec

As described in previous articles~\cite{del1,del2} the above translation is 
inadequate for odd dimensional complex representations be they for groups 
(operators in general) or for state vectors. In particular this problem 
first arose for the representation of odd dimensional spin states i.~e. 
bosons (s=0,1, ...). It is perhaps instructive to describe briefly the 
situation previous to this work for these odd dimensional states.

The first discovery was the existence of {\em anomalous} 
solutions~\cite{del3} for 
bosonic quaternionic dynamical equations (Klein-Gordon, Maxwell, ...). This 
simply followed from the complex geometry which imposes the orthogonality 
of $\psi$ and $\psi j$. If $\psi$ is chosen as the purely complex solution 
of a given bosonic equation then the purely quaternionic ($j$-$k$) solutions 
are given by $\psi j$. Or more precisely, by $j\psi$ if we wish to confront 
solutions with identical 4-momentum. This feature is already present when 
considering the eigenvalue solutions of an 
odd dimensional matrix equation. For example the (quaternionic) eigenstates 
of the 
standard spin-1 eigenvalue equation are
\bel{vec}
 \bat{1}{0}{0} \; , \; \bat{0}{1}{0} \; , \; \bat{0}{0}{1} \q \mbox{and}
\q \bat{j}{0}{0} \; , \; \bat{0}{j}{0} \; , \; \bat{0}{0}{j} \q ,
\ee
the last three (anomalous) cannot mix under rotations with the former 
because of the complex nature of the $3$-dimensional rotation generators. Thus 
the vector space is {\em reducible} but not apparently the operator 
form\f{Fermionic complex operators (but acting within a quaternionic space) 
being even-dimensional matrices are reduced to 
half the dimension with generalized quaternions, 
with the maintenance of the same number of total solutions.}. This at least 
was the belief up to know. Without the simultaneous reduction of states and 
operators it was not clear if these group representations - complex odd 
dimensional acting upon a quaternionic space - could be neglected. They 
appeared to exclude a complete translation between CQM and QQM which was  
good or bad news according to ones point of view.

There existed at least one way of bypassing this problem by eliminating the 
anomalous solutions~\cite{del4,del5}. Assuming of 
course that the anomalous solutions are an unwanted feature. This was to use 
``spinor'' (fermionic-like) equations for bosons such as the Kemmer 
equation which contains spin 1 plus spin 0 solutions. The corresponding 
Duffin-Kemmer-Petiau $\bg$-matrices (being even dimensional $16 \times 16$) 
have a reduced form of $8\times 8$ quaternionic representations. In this 
procedure anomalous solutions do not appear for the same reason that they 
do not appear for the Dirac equation. Actually the $\bg$-complex matrices 
are formed of a 10-dimensional (spin 1) and 5-dimensional (spin 0) and 
1-dimensional (unphysical $\bg=0$) block structure. Thus while the spin 1 
case is automatic because even-dimensional, the spin 0 case is handled by 
using the trick of adding 
the unphysical solution and working formally with $6\times 6$ matrices. We 
have shown elsewhere that the Kemmer quaternionic equation is thus not 
equal to the Klein-Gordon quaternionic equation etc., 
because it does not have the 
anomalous solutions. In fact it corresponds to the modified equations 
obtained by complex-projecting the original quaternionic equations.

The above trick uses explicitly the dynamical particle equations. It has 
limited the odd dimensional states to purely complex ones and thus 
apparently made the corresponding representations irreducible. {\em This is not 
the correct interpretation}. First because we must be able to discuss the 
group representations of e.~g. $SU(2, \; c)$ without any reference to 
dynamical equations. Secondly because, in fact, the odd dimensional complex 
representations {\em are} reducible with generalized quaternions thanks to 
the overlapping feature described below. Hence the problem of having a 
reducible vector space for a non-reducible matrix representation will be 
partially eliminated.

\section{Overlapping Representations}
We shall describe this technique first by deriving the situation for spin 1 
and then extracting the general rules for any odd dimensional matrix. The 
three complex antihermitian generators of spin 1 
$A_{m}$ ($m=1, \; 2, \; 3$) are in standard form
\be
A_{1}=-\frac{i}{\sqrt{2}} \; \bamt{0}{1}{0}{1}{0}{1}
{0}{1}{0} \q , \q 
A_{2}=\frac{1}{\sqrt{2}} \; \left( \ba{ccc}
0 & $-1$ & 0\\ 1 & 0 & $-1$\\ 0 & 1 & 0 \eam \q , \q 
A_{3}=-i \left( \ba{ccc}
1 & 0 & 0\\ 0 & 0 & 0\\ 0 & 0 & $-1$ \eam \q .
\ee
These have normal/anomalous solutions shown in eq.~(\ref{vec}). 

Thus each eigenvalue is degenerate and the vector space represented by the 
column matrices is reducible to two three dimensional subspaces. The 
conventional form of reductions of the $3\times 3$ generators is not 
possible because it would require the division of the $3\times 3$ matrices 
$A_{m}$ into two distinct blocks one of $2\times 2$ (quaternionic in 
general and this is possible) and one of $1\times 1$ 
(an ${\cal H}\mid {\cal C}$ number) and this can be explicitly excluded. 
Hermitian generators of $SU(2)$ (spin) within the numbers 
${\cal H}\mid {\cal C}$ 
exist, they are $i\mid i, \; j\mid i, \; k\mid i \;$\f{Note that 
$i, \; j, \; k$ are {\em antihermitian} generators and only within 
${\cal H}\mid {\cal C}$ do they have straightforward hermitian 
equivalents. With a quaternionic geometry one is restricted to only 
antihermitian generators~\cite{adl}.}. However, these 
correspond to spin 1/2 eigenvalues and not 
spin 1. It is easy to convince oneself that no one-dimensional spin 1 
representations exist in ${\cal H}\mid {\cal C}$.

This is an example of the reduction problem mentioned previously. Now we 
shall show explicitly that there exists a generalized quaternionic similarity 
matrix $S$ ($S^{\dag}=S^{-1}$) such that the $A_{m}$ are reduced to 
two overlapping $2\times 2$ block forms so that one element, the   
(2,2)-element, is common to both blocks. However if this element if not 
identically zero it is always a composite of two terms one of which 
annihilates one of the corresponding eigenvectors. Thus the two blocks may 
be unlocked and studied separately.

Explicitly an $S$ matrix such as that described above is given by:
\bml
\bea
S & = & \bamt{a}{ja}{0}{0}{0}{1}{d}{-jd}{0} \q ,\\
\nonumber  & & \\
S^{\dag} & = & \bamt{a}{0}{d}{-jd}{0}{ja}{0}{1}{0} \q ,
\eea
\eml
with
\bean
2a=1-i\mid i &\mbox{~~~~~~~extinguishes quaternionic elements} & ~~,\\
2d=1+i\mid i &\mbox{~~~~~~~extinguishes complex elements} & ~~,
\eean
whence,
\[ a^{2}+d^{2}=1 \; , \; a+d=1 \; , \;  da=ad=0 \; , \;
(ja)^{\dag}=-jd \; , \;  j a = d j \q .\]
The transformed generators $\tilde{A}_{m}=SA_{m}S^{\dag}$ are then given by
\bml
\bea
\put(82,15){\line(0,-1){47}}\put(82,15){\line(1,0){62}}  
\put(109,38){\line(0,-1){47}}\put(53,-9){\line(1,0){56}}  
\tilde{A}_{1} & = & \frac{1}{\sqrt{2}} \; \left( \ba{ccccccc}
 &     & &      & &      & \\
 &  k  & &  ka  & &  0   & \\
 &     & &      & &      & \\
 & kd  & &  0   & &  -ka & \\
 &     & &      & &      & \\
 &  0  & &  -kd & & -k   & \\
 &     & &      & &      &   \eam  \q ,\\
\put(89,15){\line(0,-1){47}}\put(89,15){\line(1,0){56}}  
\put(115,38){\line(0,-1){47}}\put(53,-9){\line(1,0){62}}  
\tilde{A}_{2} & = & \frac{1}{\sqrt{2}} \; \left( \ba{ccccccc}
        &        &   &          &  &          &   \\
        &  j     &   &  -ja     &  &  0       &   \\
        &        &   &          &  &          &   \\
        & -jd    &   &  0       &  &  ja      &   \\
        &        &   &          &  &          &   \\
        &  0     &   &  jd      &  &  -j      &   \\
        &        &   &          &  &          &   \eam  \q ,\\
\put(70,15){\line(0,-1){47}}\put(70,15){\line(1,0){40}}  
\put(91,38){\line(0,-1){47}}\put(49,-9){\line(1,0){42}}  
\tilde{A}_{3} & = & -i \; \left( \ba{ccccccc}
 &   & &  & &    &   \\
 & a & &0 & &  0 &   \\
 &   & &  & &    &   \\
 & 0 & &$-1$ & &  0 &   \\
 &   & &  & &    &   \\
 & 0 & &0 & & d  &   \\
 &   & &  & &    &   \eam  \q .
\eea
\eml
In $\tilde{A}_{3}$ the (2,2)-element can be written conveniently as~   
$i(a+d)$, i.~e. containing a sum of projection operators. The corresponding 
transformed state vectors with eigenvalues (with respect to 
$\tilde{B}_{3}\mid i$ or $\tilde{C}_{3}\mid i$) $+1, \; 0, \; -1$ are 
respectively 
\[ \bat{1}{0}{0} \; , \; \bat{j}{0}{0} \; , \; \bat{0}{1}{0} \q \mbox{and}
\q \bat{0}{0}{j} \; , \; \bat{0}{0}{1} \; , \; \bat{0}{j}{0} \q .\] 
We observe that the first triplet consists of vectors of the form
\[ \bat{q}{z}{0} \q ,\]
while the second triplet of vectors of the form
\[ \bat{0}{jz}{q} \q .\]
Naturally the two triplets remain orthogonal with a complex geometry and 
furthermore the separate sets of reduced $2\times 2$ quaternionic 
blocks  do 
not perform any rotation upon one or other set of triplets. Actually the 
two sets of reduced $2\times 2$ generators are connected by a 
similarity transformation and thus are in turn equivalent. Explicitly the 
sets of $2\times 2$ reduced quaternionic representations are:
\bc
\bt{lclclc}
$\tilde{B}_{1}=\frac{1}{\sqrt{2}}\bamd{k}{ka}{kd}{0}$ & ~~, & $\q 
\tilde{B}_{2}=\frac{-1}{\sqrt{2}}\bamd{-j}{ja}{jd}{0}$ & ~~, & $\q
\tilde{B}_{3}=-ia\left( \ba{cc} 1 & 0\\ 0 & $-1$ \eam$ & ~~;  \\
 & & & & & \\ 
$\tilde{C}_{1}=\frac{-1}{\sqrt{2}}\bamd{0}{ka}{kd}{k}$ & ~~, &  $\q
\tilde{C}_{2}=\frac{1}{\sqrt{2}}\bamd{0}{ja}{jd}{-j}$ & ~~, & $\q
\tilde{C}_{3}=-id\left( \ba{cc} $-1$ & 0\\ 0 & 1 \eam $ & ~~. 
\et
\ec
The corresponding state vectors with eigenvalues $+1, \; 0, \; -1$ respectively 
are
\[ \bad{1}{0} \; , \; \bad{j}{0} \; , \; \bad{0}{1} \q \mbox{and}
\q \bad{0}{j} \; , \; \bad{0}{1} \; , \; \bad{j}{0} \q .\]

It is of course natural to ask what the translation of these reduced 
$2\times 2$ generators to complex form yields. 
The answer, which seems obvious a 
posteriori is the $4\times 4$ complex generators of $SU(2)$ {\em reducible} 
to spin-1 $\oplus$ spin-0. Not, of course, the irreducible $4\times 4$ 
representations which would correspond to spin 3/2. Because of our derivation 
we would be tempted to identify the spin-0 element as a member of an 
independent 
triplet, but this has no physical foundation. What is significant is that 
the reduction is not perfect in the sense that it brings along a  
{\em scalar} partner.

These results lead to the following consequences: 
One is a mechanical (automatic) way of 
reducing {\em any} odd dimensional (otherwise irreducible) complex matrix 
with quaternions into {\em overlapping block} structure. The second is the 
physical significance of this procedure. For the first, we 
propose to add an extra row and column of zero's to our matrix thus making 
it become an even matrix and then applying the {\em translation} of this 
complex matrix to quaternions. This is a formal {\em trick} since we began 
with a complex odd dimensional matrix operating upon a quaternionic space 
i.~e. con\-sidered as a quaternionic matrix without need of translation and 
with only a question about its reducibility. 
Nevertheless, this trick always yields 
one or other of the overlapping block forms. For the spin-1 case 
studied above this procedure gives e.~g.
\bml
\bea
-i \left( \ba{ccc} 1 & 0 & 0\\ 0 & 0 & 0\\ 0 & 0 & $-1$ \eam \q \lraw \q & 
-i \bamq{1}{0}{0}{0} 0 & 0 & 0 & 0\\
0 & 0 & $-1$ & 0\\
0 & 0 & 0 & 0
\eam \q \lRaw \q & ia \left( \ba{cc}
$-1$ & 0\\ 0 & 1 \eam \q ,
\eea
or/and,
\bea 
-i \left( \ba{ccc} 1 & 0 & 0\\ 0 & 0 & 0\\ 0 & 0 & $-1$ \eam \q \lraw \q & 
-i \bamq{0}{0}{0}{0} 0 & 1 & 0 & 0\\
0 & 0 & 0 & 0 \\
0 & 0 & 0 & $-1$
\eam \q \lRaw \q & id \left( \ba{cc} 1 & 0\\ 0 & $-1$ \eam \q .
\eea
\eml
This procedure avoids the need of determining $S$ explicitly each time and 
is very simple. We also note that the resulting $2\times 2$ generators 
exactly reproduce the tensor product of the generators of spin 
$\frac{1}{2}\otimes \frac{1}{2}=1\oplus 0$. This brings us to the physical 
interpretation which will be given in the conclusions.

We complete this section by giving explicitly the quick-rules for translating 
(and not now reducing) a generic odd dimensional complex matrix:
\[
M \; = \; \left( \ba{cccc} 
\ddots & \vdots & \vdots & \vdots\\
\cdots & c_{1} & c_{2} & c_{3} \\
\cdots & c_{4} & c_{5} & c_{6} \\
\cdots & c_{7} & c_{8} & c_{9} \eam \q .\]
As suggested above, we add an extra row and column of zero's to our odd 
dimensional complex matrix
\[
M \; \raw \; \left( \ba{ccccc} 
\ddots & \vdots & \vdots & \vdots & \vdots \\
\cdots & c_{1} & c_{2} & c_{3} & 0  \\
\cdots & c_{4} & c_{5} & c_{6} & 0  \\
\cdots & c_{7} & c_{8} & c_{9} & 0 \\
\cdots &   0   &   0   &   0   & 0 \eam \q ,
\]
after that we can immediately translate the blocks
\[
\bamd{c_{1}}{c_{2}}{c_{4}}{c_{5}} \q , \q 
\bamd{c_{3}}{0}{c_{6}}{0} \q , \q 
\bamd{c_{7}}{c_{8}}{0}{0} \q , \q 
\bamd{c_{9}}{0}{0}{0} \q , \q ... 
\]
by the standard rules for even dimensional translation~(\ref{2a}). The 
final result brings to
\bea
M=\left( \ba{cccc} 
\ddots & \vdots & \vdots & \vdots\\
\cdots & c_{1} & c_{2} & c_{3} \\
\cdots & c_{4} & c_{5} & c_{6} \\
\cdots & c_{7} & c_{8} & c_{9} \eam 
& \q \lRaw \q & M=\bamt{\ddots}{\vdots}{\vdots}{\cdots}{Q_{1}}{Q_{2}}
{\cdots}{Q_{3}}{Q_{4}} \q ,
\eea
\bean
2Q_{1} & = & c_{1}+c_{5}^{*}+j(c_{4}-c_{2}^{*}) -i 
[c_{1}-c_{5}^{*}-j(c_{4}+c_{2}^{*})]\mid i \q ,\\
2Q_{2} & = & (c_{3}+jc_{6}) (1-i\mid i) \q ,\\
2Q_{3} & = & (1-i\mid i) (c_{7}-jc_{8}^{*})  \q ,\\
2Q_{4} & = & c_{9} (1-i\mid i) \q .
\eean

\section{Conclusions}

We began by admitting embarrassment with purely complex odd dimensional 
matrix representations of a group (our example was $SU(2)\sim U(1, \; q)$) 
acting upon a quaternionic space. The space representation was 
reducible, the generators (and 
hence group) representations were not. Furthermore, we could not translate  
odd dimensional complex vector spaces in CQM as was possible for even 
dimensional 
spaces. Now we suggest a cure for both these problems. First the odd 
dimensional complex matrices (within quaternionic space) are reducible if 
we allow for overlapping 
block structures. A {\em remnant} of say the anomalous solutions survives in 
the form of a singlet (scalar) state. We 
have suggest the procedure of adding an extra null line and column and then 
formally translating as a rapid way of deriving the reduced overlapping 
blocks. Even from this viewpoint the extra lines correspond to a 
one-dimensional zero generator (the added corner element) which implies the 
existence of an additional scalar. In the case of translations of odd 
dimensional operators from CQM to QQM we repeat this procedure of adding an 
extra null line and column and then translating. The complex state vectors 
will, of course, also acquire an extra element. This again corresponds to 
adding by hand an extra scalar particle. Whether in a given theory this 
particle is physical or not is beyond the limits of this paper. It depends 
upon the dynamics of the situation e.~g. recall the case of the Kemmer 
equation.

However, one cannot elude the impression that quaternions 
invite {\em even numbers 
of particle states}. Thus for spin 1 we are obliged to add a scalar state 
even if we avoid the full number of anomalous copies. Perhaps it is only an 
accident that at the fundamental level the number of Higgs (before 
spontaneous breaking) is four as are the number of gauge particles as are 
the number of particle (antiparticle) fermions per family (at least of left 
handed nature). We thus have no difficulty in dealing and translating the 
Salam-Weinberg model~\cite{del6}. 
We apparently have difficulty with color (for $SU(3)$ 
triplet states) but here we have two possibilities. If we include other 
multiplicative groups such as $SU(2)_{weak}$ then the overall dimensions may 
remain even and be translated as above. In this case the only price we pay is 
our inability to assign a {\em natural} quaternionic group to 
$SU(3)_{color}$. Alternatively, and far more ambitions, we would be 
tempted to use $4\times 4$ (reducible) representations of 
$SU(3)_{color}$ before translating. This corresponds to passing to 
$SU(4)_{color} \sim SU(2, \; q_{c})$ or in terms of its  
$SU(3)_{color}$ subgroup to consider 3 primary colors plus white 
(color singlet). In this case we must identify the {\em white} quark with a 
physical particle, for example the neutrino as in the Pati-Salam 
model~\cite{pat}, 
or else seek refuge in spontaneous symmetry breaking to take the white 
quark mass to values beyond present experimental limits. 

Finally we wish to remember that within quaternionic quantum mechanics with 
{\em quaternionic geometry} there is a stimulating possibility to look at 
fundamental physics as proposed by Adler~\cite{adl,adlart}. He suggest that 
the color degree of freedom postulated in the Harari-Shupe scheme~\cite{har} 
could  be sought in a {\em noncommutative} extension of standard quantum 
mechanics.

\bref
\bi{hor}
L.~P.~Horwitz and L.~C.~Biedenharn, \axp{157}{432}{84}. 
\bi{rem}
J.~Rembieli\'nski, \jxg{11}{2323}{78}.
\bi{del1}
S.~De Leo and P.~Rotelli, \nxd{B110}{33}{95}.
\bi{rot}
P.~Rotelli, \mxb{4}{933}{89}. 
\bi{del2}
S.~De Leo and P.~Rotelli, \pxxa{92}{917}{94}.
\bi{del3}
S.~De Leo and P.~Rotelli, \pxf{45}{575}{92}.
\bi{del4}
S.~De Leo and P.~Rotelli, {\it Quaternion Higgs and the Electroweak Gauge 
Group}, Int.~J.~Mod.~Phys.~A (to be published).
\bi{del5}
S.~De Leo, {\it Duffin-Kemmer-Petiau Equation on the Quaternion Field}, 
Prog.~Theor.~Phys. (to be published).
\bi{adl}
S.~L.~Adler, {\it Quaternionic Quantum Mechanics and Quantum Fields} 
(Oxford, New York, 1995).
\bi{del6}
S.~De Leo and P.~Rotelli, {\it Quaternionic Electroweak Theory}, 
(submitted for publication).
\bi{pat}
J.~C.~Pati and A.~Salam, \pxf{8}{1240}{73}.
\bi{adlart}
S.~L.~Adler, Phys.~Lett. {\bf 332B}, 358 (1994).
\bi{har}
H.~Harari, Phys.~Lett. {\bf 86B}, 83 (1979); M.~A.~Shupe, 
Phys.~Lett. {\bf 86B}, 87 (1979).

\eref

\end{document}